\documentclass{ieeeaccess}
\usepackage{cite}
\usepackage{amsmath,amssymb,amsfonts}
\usepackage{algorithmic}
\usepackage{graphicx}
\usepackage{textcomp}
\usepackage{hyperref}

\def\BibTeX{{\rm B\kern-.05em{\sc i\kern-.025em b}\kern-.08em
    T\kern-.1667em\lower.7ex\hbox{E}\kern-.125emX}}
\begin{document}
\history{Date of publication xxxx 00, 0000, date of current version xxxx 00, 0000.}
\doi{10.1109/ACCESS.2017.DOI}

\newcommand{\red}[1]{{\textcolor{black}{#1}}}

\title{Modelling strong control measures for epidemic propagation with networks --- A COVID-19 case study}
\author{\uppercase{Michael Small}\authorrefmark{1}, \IEEEmembership{Senior Member, IEEE} and
\uppercase{David Cavanagh}\authorrefmark{2}, \IEEEmembership{Member, IEEE}}
\address[1]{Complex Systems Group, Department of Mathematics and Statistics, University of Western Australia, Crawley, Perth, Western Australia (email: michael.small@uwa.edu.au)}
\address[1]{Mineral Resources, Commonwealth Scientific and Industrial Research Organisation, Kensington, Perth, Western Australia}
\address[1]{Integrated Energy Pty Ltd, Como, Perth, Western Australia (email: michael.small@integratedenergy.com.au)}
\address[2]{Integrated Energy Pty Ltd, Como, Perth, Western Australia (email: david.cavanagh@integratedenergy.com.au)}

\tfootnote{MS is supported by Australian Research Council Discovery Grants DP180100718 and DP200102961.}

\markboth
{Author \headeretal: Preparation of Papers for IEEE TRANSACTIONS and JOURNALS}
{Author \headeretal: Preparation of Papers for IEEE TRANSACTIONS and JOURNALS}

\corresp{Corresponding author: Michael Small (michael.small@uwa.edu.au).}

\begin{abstract}
 We show that precise knowledge of epidemic transmission parameters is not required to build an informative model of the spread of disease. We propose a \red{detailed model of the topology of the contact network under various external control regimes} and demonstrate that this is sufficient to capture the salient dynamical characteristics and to inform decisions. \red{Contact between individuals in the community is characterised by a contact graph, the structure of that contact graph is selected to mimic community control measures.} Our model of city-level transmission of an infectious agent (SEIR model) characterises spread via a (a) scale-free contact network (no control); (b) a random graph \red{(elimination of mass gatherings)}; and (c) small world lattice (partial to full lockdown --- ``social'' distancing). This model exhibits good \red{qualitative} agreement between simulation and data from the 2020 pandemic spread of \red{a novel coronavirus. Estimates} of the relevant rate parameters of the SEIR model are obtained and we demonstrate the robustness of our model predictions under uncertainty of those estimates.  \red{The} social context and utility of this work is identified\red{,  contributing to a highly effective pandemic response in Western Australia}.
\end{abstract}


\titlepgskip=-15pt

\maketitle

\section{Introduction}
\label{sec:introduction}

\PARstart{M}{odelling} of disease transmission via compartmental models is well established and generally \red{highly} effective. However, the differential equations \red{of these models} depend on good estimates of underlying rate parameters and \red{will then} provide a continuous solution under the assumption that the population is well-mixed and homogeneous (i.e. all individuals have equal contact with all others). Under these assumptions disease propagation is driven \red{by the parameter $R_0$} --- the ratio of the rate of new infections to the rate of removal of infectious individuals from the transmission pool. \red{Typically, and particularly for contemporary and evolving transmission, these parameters can be somewhat difficult to estimate \cite{jM20,hW20}.}

We propose an alternative approach to modelling the dynamic transmission of diseases. \red{A consequence of this alternative approach is that the main determinant of epidemic dynamic behaviour is the contact network between individuals rather than precisely chosen \em{optimal} values of epidemic rate parameters.} The physics literature is rife with \red{ models of propagation dynamics} on networks. We observe that different societal control measures manifest as distinct topological structures and model city-level transmission of an infectious agent. Our approach models changing control strategies by changing the features of the underlying contact network with time. This approach allows us to model the likely time course of a disease and, perhaps surprisingly, we find that \red{this} approaches \red{is} both quantifiable and robust to uncertainty of the underlying rate parameters.

This report is intended as a guide to computational modelling of reported epidemic infection rates when good estimates of underlying epidemiological rate parameters are not available. The model provides a useful prediction of current control strategies. Nonetheless, we emphasise that the methodology and techniques are not (of themselves) novel, they have been discussed extensively in the references cited herein. \red{The primary novel contribution of this paper is the interpretation of complex network topologies as the principal relevant parameter to characterise control, and a commentary on the live application of this approach} in pandemic response and recovery. \red{While global efforts to model the spread and control of coronavirus continue, we are taking a decidedly local approach. We focus our modelling and discussion on transmission in Australian cities, which we characterise as large heterogeneous populations. In particular, we focus on the most isolated of all cities in Australia, Perth, the capital of Western Australia. For the purposes of this manuscript we treat Perth as a single isolated urban centre of approximately 2 million people. Epidemic parameters, which we describe later, either follow established epidemiological values or are estimated to fit the time course of infection data. While our model is specific to one city, we intend that the methods and conclusion are generic and will be useful elsewhere.}

\red{Our model is a model of contact graphs. Different contact graphs are utilised to model different contact patterns within the population and hence model the effect of different control measures. In Sec. \ref{intro} we introduce and discuss a small amount of the most  relevant literature, and in Sec. \ref{model} we proceed to describe our model.}  

\section{Background}
\label{intro}

It is no exaggeration to say that pandemic spread of infectious agents has very recently attracted wide interest. Mathematical epidemiology is a venerable and well respected  field \cite{jM93}. Propagation of disease in a community is modelled, under the assumption of a well-mixed and homogeneous population via differential equations characterising movement of individuals between disease classes: susceptible (S), exposed (E), infectious (I) or removed (R). 

The standard compartmental (i.e. SIR) model dates back to the mathematical {\em tour de force}  of Kermack and McKendrick \cite{wK27}. The model assumes individuals can be categorised into one of several {\em compartments}: S or I; S, I or R; or, S, E, I, or R being the most common. Transition between the various compartments is governed by rate parameters $a$ and $r$ and it is the job of the mathematical epidemiologist to estimate those rates --- and hence, when $\frac{dI}{dt}<0$ and the transmission is under control. In the standard SIR formulation the condition $\frac{dI}{dt}<0$ can be expressed as $\frac{aI(t)}{r}\equiv R_0<1$. Somewhat confusingly, $R_0$ is also used in the physics literature to denote the threshold itself --- as in \cite{rP01} where $R_0$ is derived in terms of moments of the contact network degree distribution. Nonetheless, efforts to estimate the relevant parameters for the coronavirus pandemic are currently underway and are best summarised (from our local perspective) by the technical reports of Shearer\cite{fS20}, Moss\cite{fM20} and co-workers.

Conversely, \red{ the renaissance of interest in mathematical graphs under the guise of complex networks \cite{aB99,pE59},  has raised considerable interest} in propagation of infectious agent-like dynamics on such structures \red{\cite{li2020,li2019}}. Commonly, the agent is either modelling the spread of information or infection. When one is restricted to the spread of infectious agents on a network (in the context of epidemiology, a contact graph) several interesting features arise. In particular, if that contact graph is a scale free network (i.e. it has a power-law degree distribution), then the criterion on the key epidemic threshold to ensure control of the outbreak (for the SIS model) becomes $R_0= 0$. \cite{rP01} \red{Disease transmission becomes faster than exponential.} In effect, what happens is that the power-law distribution of the scale-free network ensures that there is finite probability of the epidemic reaching an individual with an arbitrary large number of contacts. The number of secondary infections arising from that individual will be unbounded and transmission is guaranteed to persist. Of course, in the real-world nothing is unbounded and Fu and co-workers \cite{xF08} showed that a piece-wise linear/constant infectivity was enough to ensure a positive epidemic threshold.

Surprisingly, however, little of the work in the physics literature on epidemic transmission has examined transmission on real-world networks. The first evidence (to the best of our knowledge) that epidemic transmission did really occur on a scale-free contact graph was provided by Small and others \cite{birdflu} for the transmission of avian influenza in migratory bird populations. Curiously, though, the data presented there gave an exponent for the scale-free distribution of approximately $1.2$, significantly lower than \red{the often cited  ``usual''} range of $(2,3)$ --- that is, the distribution not only had divergent variance, but also divergent mean.

The emergence of an earlier coronavirus, associated with the Severe Acute Respiratory Syndrome (SARS) \red{in 2003, provided an opportunity} to apply the structures and concepts of complex systems to the modelling of infectious diseases. Small and Tse\cite{sars4}  introduced a complex network based model of propagation and showed good agreement between simulations of that model and available case data. \red{They} found that epidemic parameters widely quoted in the literature were only consistent with observed case data when including significant nosocomial transmission.\cite{sars3} Finally, and most importantly for the current discussion, the scale-free topology of the \red{model \cite{sars4}  explained super-spreader events through contact rather than requiring pathologically highly infectious individuals.\cite{sars6}}

\red{Both the network-based models used to model SARS in 2003 \cite{sars3,sars4,sars6}, and the model we describe here are network models of contact between individuals.} Unlike what we will propose in this current communication, \red{the model of SARS in 2003 was topologically stationary \cite{sars3,sars4,sars6}.} The model assumed a lattice with long-range (i.e. small-world) connections following a power-law degree distribution. In those papers\cite{sars3,sars4,sars6} time varying \red{control strategies were} reflected only in changes of the rate parameters. The current coronovirus outbreak (that is, COVID-19) poses a different and unique challenge. Since February 2020 (and up to the time of writing) global transport networks and daily movement of individuals have been disrupted on a global scale. Entire cities and countries have engaged in various levels of ``lockdown''. We argue that it is \red{neither appropriate nor sufficient} to model this simply by modifying the rate of transmission or rate of removal. \red{In the current work we propose a network switching model through which the topology of the network changes to reflect various changes in government and community mitigation and control strategies.}

In Sec. \ref{model} we introduce our model structure, and Sec. \ref{growthsec}  explores analytic expressions for the epidemic growth rate. In Sec. \ref{results} presents our results for the case study of most interest to us, and Sec. \ref{paramfit} describes a process for optimising model parameters based on observed caseload data. 

\section{The model}
\label{model}

We assume nodes in our network can be in one of four states, corresponding to the four states of the standard SEIR model: susceptible $S$, exposed $E$, infected $I$, and removed $R$. Rates govern the probability of transition between these states, with transition from $S$ to $E$ \red{occurring only through neighbour-to-neighbour contact on the graph with a node in state $I$.}

\red{For} comparison, the standard SEIR compartmental differential equation based formulation is given by \cite{jM93}:
\begin{eqnarray}
\label{seir}
\left(\begin{array}{c}
\frac{dS}{dt}\\
\frac{dE}{dt}\\
\frac{dI}{dt}\\
\frac{dR}{dt}
\end{array} 
\right)
\equiv
\left(\begin{array}{c}
S'\\
E'\\
I'\\
R'
\end{array} 
\right)
& = &
\left(
\begin{array}{c}
-pSI\\
+pSI - qE\\
+qE - rI\\
rI
\end{array}\right),
\end{eqnarray}
where  \red{$p,q,r\in (0,1]$} determine the rate of infection, latency and removal respectively. Clearly, if we desire $\frac{dE}{dt}+\frac{dI}{dt}<0$ we need  $\frac{pS(t)}{r}<1$. The parameter $q$ determines the average latency period, and hence the ratio of $p$ and $r$ determines the rate of spread.

The assumptions underpinning models of the form (\ref{seir}) are that the population is fully mixed --- that is, contacts exist between all members of the population, or, rather, every individual is indistinguishable and contacts occur at a constant rate between individuals. One way to extend this model \red{is to introduce multi-group or stratified (perhaps by age, vulnerability, comorbidity, or location) transmission models. }Doing so for \red{coronavirus transmission} is reasonable and has been extensively covered elsewhere: for the Australian perspective see \cite{gM20,sC20} However, this would require estimating distinct values of $p$, $q$ and $r$ for each strata. \red{We choose an approach which has fewer free parameters}\footnote{Our model is a network. One could argue that a network of $N$ individuals has $N(N-1)/2$ parameters $p_{ij}$ governing contact between individual $i$ and $j$ and (worse) potentially unique $r_i$ and $q_i$ for each individual. We prefer the statistical physics approach of describing the key features of a network with a very small number of parameters.} \red{and model infection at the daily time scale.} 

 \begin{figure*}[t]
\centering
\includegraphics[width=\linewidth]{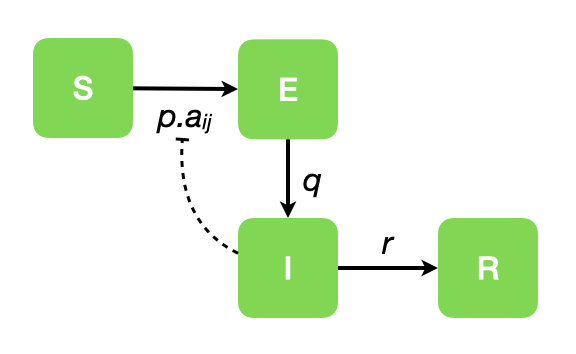}
\caption{\red{Model flow chart. A Graphical representation of the model state transition process. Each node can be in one of four states $S$, $E$, $I$, or $R$ with transition between them determined by probabilities $p$, $q$ and $r$ and the contact process of elements $a_{ij}$ of the network adjacency matrix $A$. Hence node-$i$ has probability $pa_{ij}$ of being infected through contact with node-$j$.}}
\label{flowchart}
\end{figure*}

Let $A$ be an $N$-by-$N$ binary symmetric matrix, $a_{ij}=1$ iff \red{individuals} $i$ and $j$ are in contact. \red{The matrix $A$ is the adjacency matrix of the contact network which we model.} We suppose that all individuals, excluding a small number \red{who} are exposed (E), are initially susceptible (S). Then, at each time step \red{(each day)}:
\begin{itemize}
\item[$S\rightarrow E$] a susceptible node $i$ becomes exposed if there exists a node $j$ that is infectious (I) and $a_{ij}=1$  with probability $p$;
\item[$E\rightarrow I$] an exposed node becomes infectious with probability $q$; and,
\item[$I\rightarrow R$] an infectious node becomes removed (R) with probability $r$. 
\end{itemize}
\red{The model structure is depicted in Fig. \ref{flowchart}.} Structural patterns of contact within the community are then modelled by varying the structure of the network $A$. \red{In this paper we propose distinct models corresponding to the different control strategies. in the following four subsections, the control strategies which we consider are: \ref{B} no control, modelled  with a scale free network; \ref{L0} hard isolation, modelled as a lattice; \ref{L1} no mass gatherings via a random graph; and \ref{Ls} ``social'' distancing via a small-world network. We explore these four distinct network structures in the following subsections.}

\subsection{Scale-free network $B$} 
\label{B}
Let $B$ denote an $N$-by-$N$ unweighted and undirected scale-free network. For simplicity (and rapidity of calculation) we generate this network via the preferential attachment algorithm of Barabasi and Albert \cite{aB99} --- there are good reasons for not doing this (\red{notably that the rich club will be highly connected} \cite{linjun1,growing,prefpa}). Nonetheless, simulations presented here did not depend on the choice of the Barabasi-Albert model over alternatives including the configuration model or likelihood approaches \cite{prefpa}. The network $B$ is parameterised by \red{$\frac{k}{2}$ the number of new edges associated with each new node and so we represent it as $B(k)$ (if each new node contributes $\frac{k}{2}$ new edges, then the mean degree will be $k$). Here, to ensure comparable number of edges, we choose $k=4$.}

The network $B(k)$ provides a model of random contacts in a community. There is ample evidence that individual contact patterns follow an approximately scale-free distribution. Specifically, in the context of the current pandemic, there is clear evidence in large scale community spread of COVID-19 at sporting events and other mass gatherings which are well modelled via the tail of a scale-free distribution \red{\cite{aE20,dM20,dA20}}. Due to the random wiring of connections between nodes we expect contact network $B$ to yield \red{at least} exponential growth of infection. \red{The tail of the degree distribution is unbounded and so the actual growth rate is greater than exponential.}

\subsection{Regular lattice $L(0)$}
\label{L0}
Let $L$ denote a regular two dimensional lattice with periodic boundary conditions. Each node has four adjacent neighbours. For consistency with what follows we denote this as $L(0)$. Growth of infection on a lattice will be equivalent to diffusion in two dimensions and hence the infected population will grow geometrically -- in the case of the configuration discussed here growth is sub-linear. 

Lattice configuration is used here as an approximation to {\em hard isolation}: individuals do not move in geographical space and are therefore only connected to neighbours. Intuitively, one might expect a hard isolation model to consist of small isolated clusters corresponding to individual family units. In addition to being uninteresting -- for the very obvious reason that transmission would cease -- such a model is overly optimistic. Transmission would still be expected to occur between neighbours (in the ordinary sense of the word). The regular lattice configuration model is able to model such infection between family units, and adjacent dwellings. \red{This is exactly the philosophy behind the model structure of \cite{sars3}.}

\subsection{Random Graph $L(1)$}
\label{L1}
Let $L(1)$ denote a random graph (ala Erd\"os-Renyi \cite{pE59}) with \red{mean degree equal to four.} Connections between nodes are chosen uniformly at random and constrained to avoid multiple edges or self-loops. \red{That is, each edge is  assigned to \red{connect} two randomly chosen nodes within the network, subject to the constraint of no self-loops and no multiple edges.} At the opposite extreme to $L(0)$ we denote by $L(1)$ the lattice graph with no lattice structure --- all connections have been rewired and hence correspond to complete random wiring. In other words, while $L(1)$ is not a lattice it is the limiting case of $L(q)$ for $q\rightarrow 1$. Unlike $B$ the degree distribution of $L(1)$ is binomial \red{(there is a fixed constant probability that a link exists between any two random nodes, independent of all other structure)}. Hence, while $B$ will be characterised by super-spreader events (spiky outliers in the daily infection count), \red{spreading with $L(1)$ contacts} is exponential but devoid of extreme events.

The random graph model represents a mixing populace with limitations placed on mass gatherings.

\subsection{Small-World lattice $L(s)$}
\label{Ls}
Finally, let $L(s)$ denote a Watts-Strogatz \cite{dW98} {\em two}-dimensional lattice with random rewiring with probability $s$. That is, the network $L(s)$ is constructed as a regular lattice $L(0)$ each edge emanating from node-$i$ has a probability $s$ of being disconnected from the neighbour node-$j$ and then rewired between node $i$ and random node-$k$ (in doing so, one node will decrease in degree by one, and one will increase by one).

For $s>0$, the graph $L(s)$ is an imperfect approximation to $L(0)$. That is, individuals are bound in a lattice configuration due to being geographically constrained. However, a fraction of individuals still exhibit long range connections. Effectively, the model $L(s)$ assumes that the populace is practising what is referred to in the popular press as ``social distancing'' (everyone is fixed at a home location and connected only to others in the same vicinity). However, there is some finite limit to compliance with the enforced isolation. A probability $s$ of a given link switching and therefore connecting random nodes corresponds to a fraction
 $c=(1-s)^k$ of nodes compliant with these distancing measures since all there $k$ edges are not switched.
 
 In opposition to the standard and rather flawed nomenclature, we will refer to this control strategy as {\em physical distancing}.
 
 \section{\red{Growth rates}}
\label{growthsec}

 We now provide estimates of the characteristic growth rates for propagation on the network structures described above. \red{A widely used} approach \cite{disbook} is to replace the compartmental \red{equations (\ref{seir})} with distinct equations for nodes of each degree. \red{What we describe here is the approach commonly adopted in the physics literature. For an excellent treatment of the original theoretical biology approach see \cite{adO08}.} Let $S_k$, $E_k$, $I_k$ and $R_k$ denote the number of nodes of degree $k$ in state $S$, $E$, $I$ or $R$. The system (\ref{seir}) then becomes
 \begin{eqnarray}
\label{seirk}
\left(\begin{array}{c}
S'_k\\
E'_k\\
I'_k\\
R'_k
\end{array} 
\right)_k
& = &
\left(
\begin{array}{c}
-pS_k\sum_\ell kP(\ell)I_\ell\\
pS_k\sum_\ell kP(\ell)I_\ell - qE_k\\
qE_k - rI_k\\
rI_k
\end{array}\right)_k,
\end{eqnarray}
 where $P(\ell)$ is the degree distribution of the network $A$. In general $P(\ell)$ is a little unsatisfactory as we should really compute the sum over $P(\ell|k)$. But even in the SIS case, doing so becomes rather unwieldy. \red{Conversely, for SEIR-type or (SIR) epidemics} the asymptotic state is trivial: $S_k(t)\rightarrow S_k^*\in(0,S(0))$, $I_k(t),E_k(t)\rightarrow 0$ and $R_k(t)\rightarrow R_k^*\in(0,S(0))$. This provides no insight.
 
Nonetheless, we are interested in growth rate which is determined via decrease in the susceptible population 
\[pkS_k\sum_\ell P(\ell)I_\ell.\] In the scale free case $P(\ell)\propto\ell^{-\gamma}$ and hence growth is super-exponential: high degree nodes have a contact rate proportional to their degree and a non-zero probability of connecting to other high degree nodes. 

Conversely, suppose that each node has a fixed degree 
\begin{eqnarray*}
P(\ell)=\left\{\begin{array}{cc}
1 & \ell=L\\
0 & {\rm{otherwise}}
\end{array}\right..
\end{eqnarray*}
In our lattice model $L=4$. The growth rate is then given by $pS_kLI_L$, system (\ref{seirk}) immediately reduces to (\ref{seir}), and one is left with the usual exponential growth or decay. Hereafter, we are considering only nodes of degree $k=L$ and will drop the subscript $k$ for convenience. However, for $s<1$ this reasoning is flawed.

Employing (\ref{seir}), assumes perfect mixing and hence random distribution of infectious and susceptible nodes on the lattice. Under diffusion the infectious nodes will spread in a single cluster: nodes in that cluster will be in class E, I or R and the remainder of the population will be susceptible. The cluster will be of size $E+I+R$ and the exposed boundary will be of size scaling with $\sqrt{E+I+R}$, nodes on that exterior will be either $E$ or $I$ (we assume that diffusion is fast enough that the removed nodes are interior ---  this is certainly only an approximation and will depend on relative values of $p$, $q$ and $r$), but only the nodes in state $I$ are infectious. Hence, the number of infectious nodes in contact with susceptibles will scale with a quantity bounded by $\frac{I}{E+I+R}\sqrt{E+I+R}$ and $\sqrt{E+I+R}$ --- mostly likely around $\frac{I}{E+I}\sqrt{E+I+R}$.\footnote{Throughout, we've assumed a 2-D lattice. Of course, this choice is arbitrary and an $N$-D lattice would naturally lead to an expression involve the exponent ${N-1}{N}$ with the random model prevailing at $n\rightarrow\infty$.} On average, only half the links from an infectious node will point to a susceptible (the remainder will point to other nodes in the cluster), hence, the number of susceptible nodes connected to an infected node is approximated by \red{\[\frac{I}{2(E+I)}\sqrt{E+I+R}\]} and the proportion of susceptible nodes that satisfy this condition will be 
\begin{eqnarray*}
\frac{I\sqrt{E+I+R}}{2S(E+I)}.
\end{eqnarray*}
\red{Hence,} the expected number of new infections from a lattice diffusion model is obtained from the product of the rate $p$ and the contact between these exposed infected and susceptible individuals
\begin{eqnarray*}
\red{p\times\left(\frac{I\sqrt{E+I+R}}{2(E+I)}\right)\times\left(\frac{I}{2(E+I)}\sqrt{E+I+R}\right)}\\
\red{= p\times \left(\frac{I^2(E+I+R)}{4(E+I)^2}\right).}
\end{eqnarray*}
We note in passing that typically $S\gg E\propto I>R$ --- certainly during initial growth, or in the case of limited penetration. Moreover, the arguments above only hold when $S\gg E,I$.

Finally, in a small-world model there is probability $s$ of a link pointing to a random distant location. With $S\gg E+I+R$ we assume that that link is pointing to a susceptible node and so the expected number of new infections is now 
\begin{eqnarray*}
(1-s)p\left(\frac{I^2(E+I+R)}{\red{4}(E+I)^2}\right)+spSI.
\end{eqnarray*}
Since, $E$ and $I$ are linearly proportionate, the first term scales (very roughly) like $(E+I+R)$ the second like $SI$. That is, a mixture of the sub-linear growth dictated by the lattice (with proportion $1-s$) and the classical compartmental model (\ref{seir}) with probability $s$. \red{Considering the $E$ and $I$ individuals as a single pool, the rate of new infections is balanced by the rate of removal $rI$ and so infection will grow if
\begin{eqnarray*}
\frac{p}{r}\left((1-s)\left(\frac{I(E+I+R)}{\red{4}(E+I)^2}\right)+sS\right) &> &1.
\end{eqnarray*}}

\section{Results}
\label{results}

 \begin{figure*}[t!]
\centering
\includegraphics[width=\linewidth]{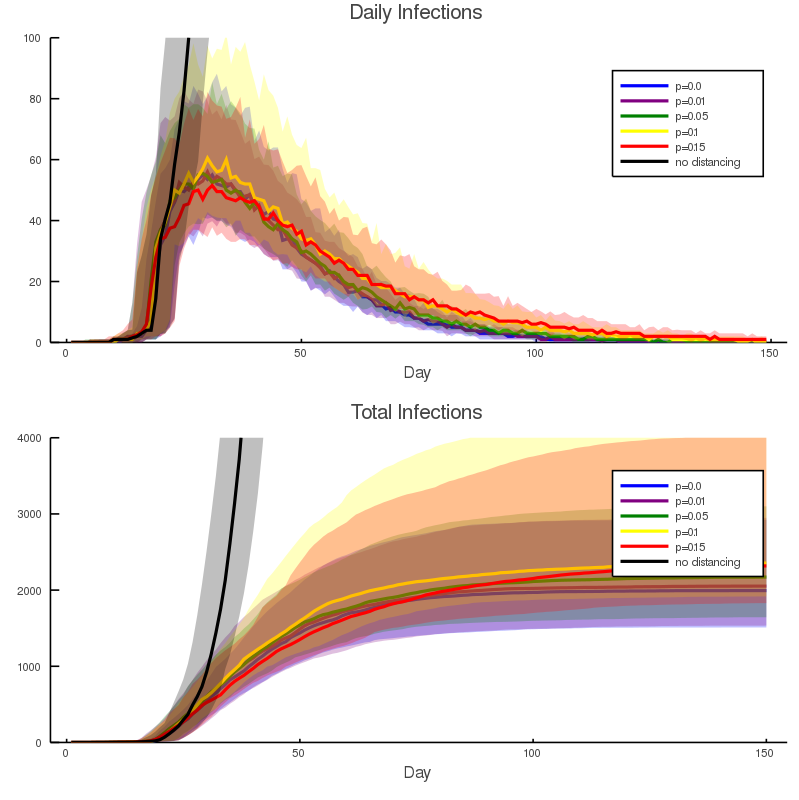}
\caption{Predicted epidemic time series. The upper panel is daily new infected individuals (i.e. $-\Delta (S(t)+E(t))$), lower panel is total number infected ($S(0)-S(t)$). For each network configuration, results show mean and distribution of $100$ simulations over $240$ days. In black $A=B(4)$ for all time. In other simulations \red{$A=B(4)$} for all $t$ until $I(t)>150$, otherwise $A=L(s)$ with values of $s$ from $0$ to $1$. The shaded envelopes are $90\%$ confidence intervals. To compute the fraction of population compliant with social isolation measures $d$ we compute $d=P({\rm no\ rewired\ links})=(1-s)^k$ (here, the number of neighbours $k=4$). \red{Epidemic parameters follow the values established for our later simulations in Table \ref{paramtable}. For the purposes of this plot, we vary only $s$ --- \red{the rewiring probability from $s=0$ to $s=0.15$}.}
}
\label{example}
\end{figure*}

\begin{table}
\begin{tabular}{l|cc}
\;\;\;\;\;\;\;\;& $t<t^*$ & $t>t*$\\\hline
 $N$ & \multicolumn{2}{c}{ $1450^2$}\\ 
 $I_\rm{th}$ & \multicolumn{2}{c}{ $150$}\\
$p$   &  $0.2$ & $\frac{1}{12}$ \\
$q$   &  $\frac{1}{7}$ & $\frac{1}{7}$\\
$r$   &  $\frac{1}{14}$ & $\frac{1}{4}$\\
\end{tabular}
\caption{Epidemic simulation parameters. The simulation size $N$ is chosen to be a square number to make the construction of $L(s)$ simpler. \red{Latency period of $q=\frac{1}{7}$ is comparable to observation, the other parameters are estimated derived from the values used in \cite{fS20,fM20} for Australian populations. These parameter values ensure growth in infection for $t<t^*$ but barely endemic otherwise (for $A \neq B$). That is, these parameters are selected to match the observed data for our principle region of interest. Subsequent parameter sensitivity computation will indicate that variation of these parameters does} not change the qualitative features, only the scale of the observed simulations.} 
\label{paramtable}
\end{table}

In this section we first present results of the application of this model. We choose a city of population of approximately $2.1\times 10^6$ (Perth, Western Australia) and perform a simulation with initial exposed seeds and contact network \red{$A=B$ (for $0\leq t<t^*$).} The transition time $t^*$ is the time with $I(t)>I_\rm{th}$ for some threshold infection load $I_\rm{th}$ for the first time (i.e. $I(t)<I_\rm{th}$ for all $t<t^*$ and $I(t^*)\geq I_\rm{th}$. For $t>t^*$ we set $B=L(s)$ for various values of $s$. \red{In what follows we will use $p(t>t^*)$ to denote the value of parameter $p$ assumed for all time $t>t^*$, similar notation is adopted for $p(t<t^*)$ and also for parameter $r$.}

The epidemic parameters which we have chosen for this simulation are illustrated in Table \ref{paramtable}.  \red{We do not wish to dwell on the epidemiological appropriateness of these parameters --- except to say that the were chosen to be consistent with our understanding of epidemiology and also gave results that appropriately coincided with the available time series data. The specific parameter values described in Table \ref{paramtable} were computed to be consistent with those employed by  \cite{fS20,fM20}. However, the models described in  \cite{fS20,fM20} are more epidemiologically detailed than ours and hence the parameter values reported here represent an agglomeration of various rates. Moreover, we confirm empirically that the rate of spread implied by these parameter choices shows very good agreement with the transmission data for Australia --- see Sec. \ref{paramest}}.

\begin{figure*}[t]
\centering
\includegraphics[width=0.8\linewidth]{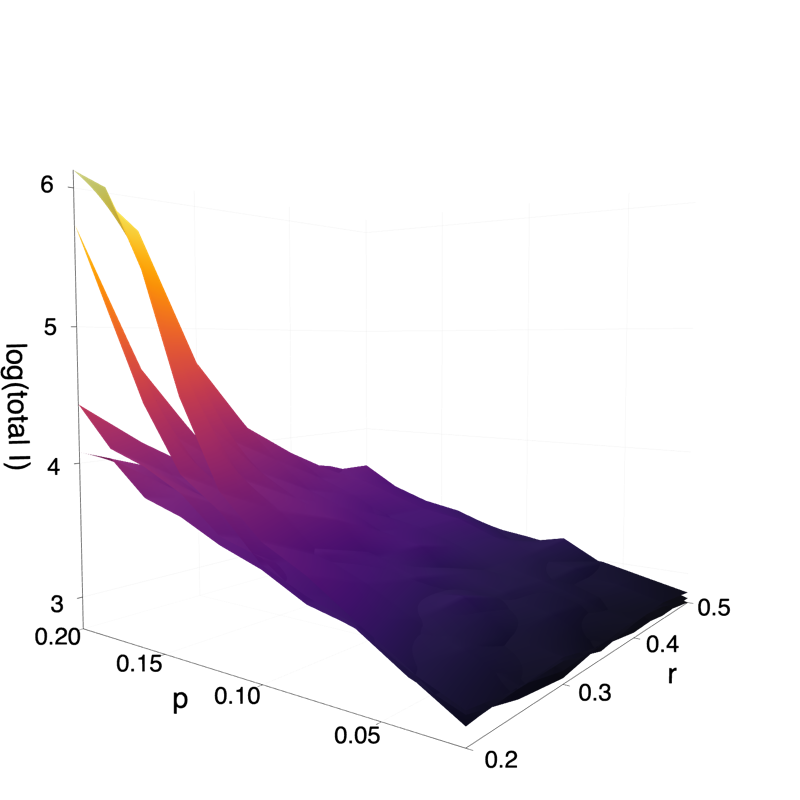}
\caption{Parameter sensitivity. The \red{four surfaces} explore the expected total number of infections (population $N=1450^2$) for various parameter values $p$ and $r$ (for $t>t^*$) and different control strategies (i.e. $L(s)$ for different $s$). The four surfaces depicted here correspond to (a) $s=0.0025$; (b) $s=0.026$; (c) $s=0.054$; (d) $s=0.065$  \red{(that is, $99\%$, $90\%$, $80\%$ and $70\%$ observance of physical distancing measures).} The three coordinates are (x) $r$; (y) $p$; and (z) $\log(\max_t (S(0)-S(t)))$ (the logarithm base-10 of the total number of infections). In each case we computed $80$ simulations of $300$ days. Other parameters are as reported in Table \ref{paramtable}. Surface (a) and (b) exhibit linear scaling with changing parameter values $p(t>t^*)$ and $r(t>t^*)$, while for (c) and (d) that growth is exponential. That is, when compliance with isolation measures drops below $90\%$ there is an explosive growth in the level of infection with $p(t>t^*)$ and $r(t>t^*)$. }
\label{parameters}
\end{figure*}

Some brief notes on the effect of parameter selection are in order. First, varying $I_\rm{th}$ will delay the transition to a ``controlled'' regime and produce a larger peak. The parameter $q$ is largely determined by the epidemiology of the infection, and for coronavirus COVID-19 \red{is fairly well established \cite{fM20}.} It does have an important influence on the time delay of the system, but that is not evident from Fig. \ref{example}. Second, the parameters $p$ and $r$ for $t<t^*$ also determine the initial rate of spread --- as standard epidemiology would expect. Third, the value of these parameters for $t>t^*$ determine the length of the ``tail''. In all simulations these parameters are chosen so that a well mixed population would sustain endemic infection. It is the network structure, not fudging of these parameters that causes extinction of the infection --- this will be further illustrated in Fig. \ref{parameters}.

Figure \ref{example} depicts one ensemble of simulations. Of note from Fig. \ref{example} is the complete infection of the population without control. Conversely, the random Erd\"os-Renyi graph $L(1)$ has a sufficiently narrow degree distribution that the infection does (slowly) die away. Various values of $L(s)$ with $s\in(0,1)$ have  the expected effect of gradually decreasing the total extent and duration of the outbreak. However, it is important to note that the $90\%$ confidence windows are very wide and overlap almost entirely --- while, on average smaller $s$ is better this is very often not evident from individual simulations. \red{This is due to random variation in the initial spread for $t<t^*$. }

\begin{figure}[t]
\centering
\includegraphics[width=\linewidth]{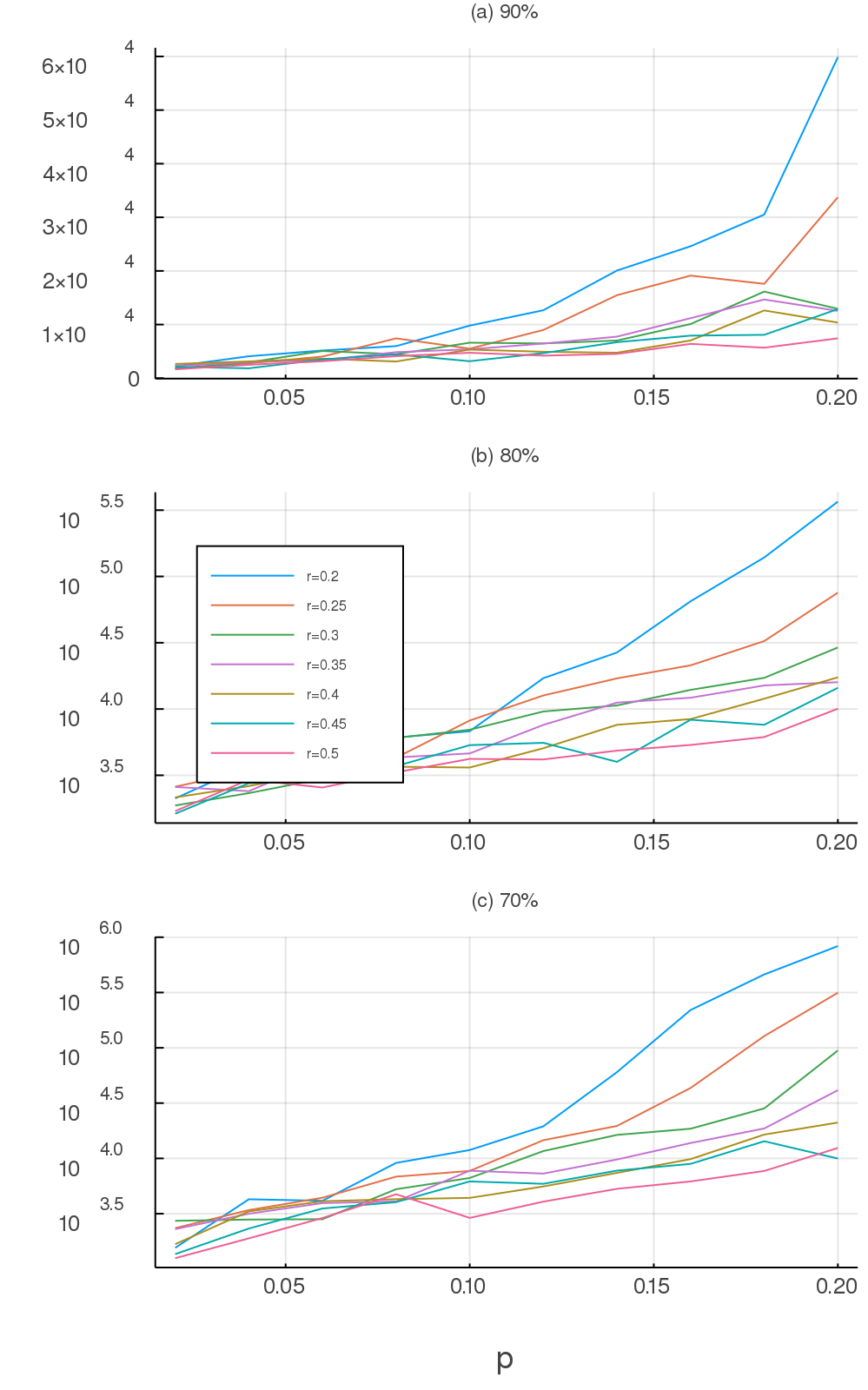}
\caption{Parameter sensitivity. The three panels explore the expected total number of infections (population $N=1450^2$) for various parameter values $p(t>t^*)$ and $r(t>t^*)$ (i.e $p$ and $r$ for $t>t^*$) and different control strategies (i.e. $L(s)$ for different $s$). The four panels depicted here correspond to (a) $s=0.013$; (b) $s=0.026$; (c) $s=0.054$ \red{($90\%$, $80\%$, $70\%$ physical distancing as reported in the panel headings)}. In each case we computed $80$ simulations of $300$ days. Other parameters are as reported in Table \ref{paramtable}). Note that panel (a) has a linear ordinate, panel (b) and (c) are depicted with a logarithmic scale. As in Fig. \ref{parameters} we observe explosive growth in impact with lower levels of compliance.
}
\label{parameters2}
\end{figure}

It is very clear from Fig. \ref{example} that the variance between simulations is similar in magnitude to variation in parameters. However, parameters in Fig. \ref{example} correspond to moderate parameters $p$ and $r$ and a wide variation in social isolation. In an effort to understand the parameter sensitivity of this simulation we perform repeated simulations over a wide range of $p(t>t^*)$ and $r(t>t^*)$. For all selected values we generate $20$ simulations of $300$ days each and compute several indicators of infection penetration
\begin{itemize}
    \item{\bf Mean total infection:} The total number of individuals that become exposed, infected or removed during the duration of the simulation. That is, $\max_t S(0)-S(t)=S(0)-S(300)$.
    \item{\bf Mean maximum infected:} The maximum daily reported number of infections - that is, the maximum number of new infected individuals: $-\max_t (S(t)+E(t)-S(t-1)-E(t-1)$
    \item{\bf Half recovered time:} The time in days required for half the simulations to entirely eliminate infection. That is, the median (over simulations) of the minimum (over time) $t$ such that $E(t)+I(t)=0$
\end{itemize}
Results for $I_\rm{th}=100$ are reported in Fig. \ref{parameters}, varying $I_\rm{th}$ simply scales the reported numbers \red{(data not shown)}. Depicted in Figs. \ref{parameters} and \ref{parameters2} are computed values of the mean total infection. The other parameters described above behave in a consistent manner.

\red{Figures \ref{parameters} and \ref{parameters2} starkly illustrate} the importance, for the coronavirus pandemic of 2020, of implementing and stringently enforcing isolation. Without isolation the epidemic impact is limited \red{only} for very optimistic values of $p(t>t^*)$ and $r(t>t^*)$. Otherwise, the mean behaviour indicates infection growth by two orders of magnitude within 300 days - almost complete penetration. Our simulations indicate that this first becomes a risk as \red{physical distancing} is less than $90\%$ effective. There is a boundary in our simulations which appears below $90\%$ isolation and grows to include even moderate values of the other epidemic control parameters $p(t>t^*)$ and $r(t>t^*)$.

\section{Parameter selection}
\label{paramfit}

In part, our aim with this communication is to dissuade the application of modelling of \red{time series} to predict certain specific futures. That is, we are interested in simulation and inferring structure from the ensemble of such simulations. The random variation reported in Fig. \ref{example} should \red{discourage} all but the most determined from prediction. Nonetheless, it is valid to ask two questions of observed time series data: (1) what parameter values are most likely given this observed trajectory, and (2) which trajectory (or set of trajectories) are most consistent with the current state. The first question we will address via a greedy optimisation procedure, to be described below. The second question is equivalent to asking for an ensemble estimate of the current state of exposed but undetected individuals within the community. A complete study of this second problem is beyond the scope of the present discussion, but some points are worth considering before we return to the issue of parameter estimation in Sec. \ref{paramest}. \red{Finally, in Sec. \ref{recovery} we provide some estimates of the effectiveness of various control measures during  recovery phase, subsequent to localised eradication.}

\subsection{State estimation}
\label{state}
As noted previously, there is very significant variation between trajectories for the same model parameter settings. While this means that the construction of more complex models -- solely from time series data -- is inadvisable, it is natural to seek to explain this variability. Simulations conducted above for an SEIR model with nontrivial latency period \red{indicates} that at any instance in time there is a large number of exposed but undetected individuals within the network. The location of this exposed class within the network (their distribution relative to hubs, for example) explains the variation we observe. This has been demonstrated by simulation from repeated random distributions of exposed individuals. It is easy to estimate the expected number $E(t)$ from the time series $I(t)$ and $R(t)$, however, the distribution of these individuals on the network is not uniform. The question that must be addressed to resolve this issue is what is the expected distribution of $E(t)$ random walkers on a network $A$? In the interest of clarity and succinctness, we do not address this issue here. \red{For the purposes of Sec. \ref{recovery}, below, we simply model a re-introduction of infection as a small number of exposed individuals randomly distributed on the contact graph.}

\subsection{Parameter estimation}
\label{paramest}

\begin{figure*}[t]
\centering
\includegraphics[width=\linewidth]{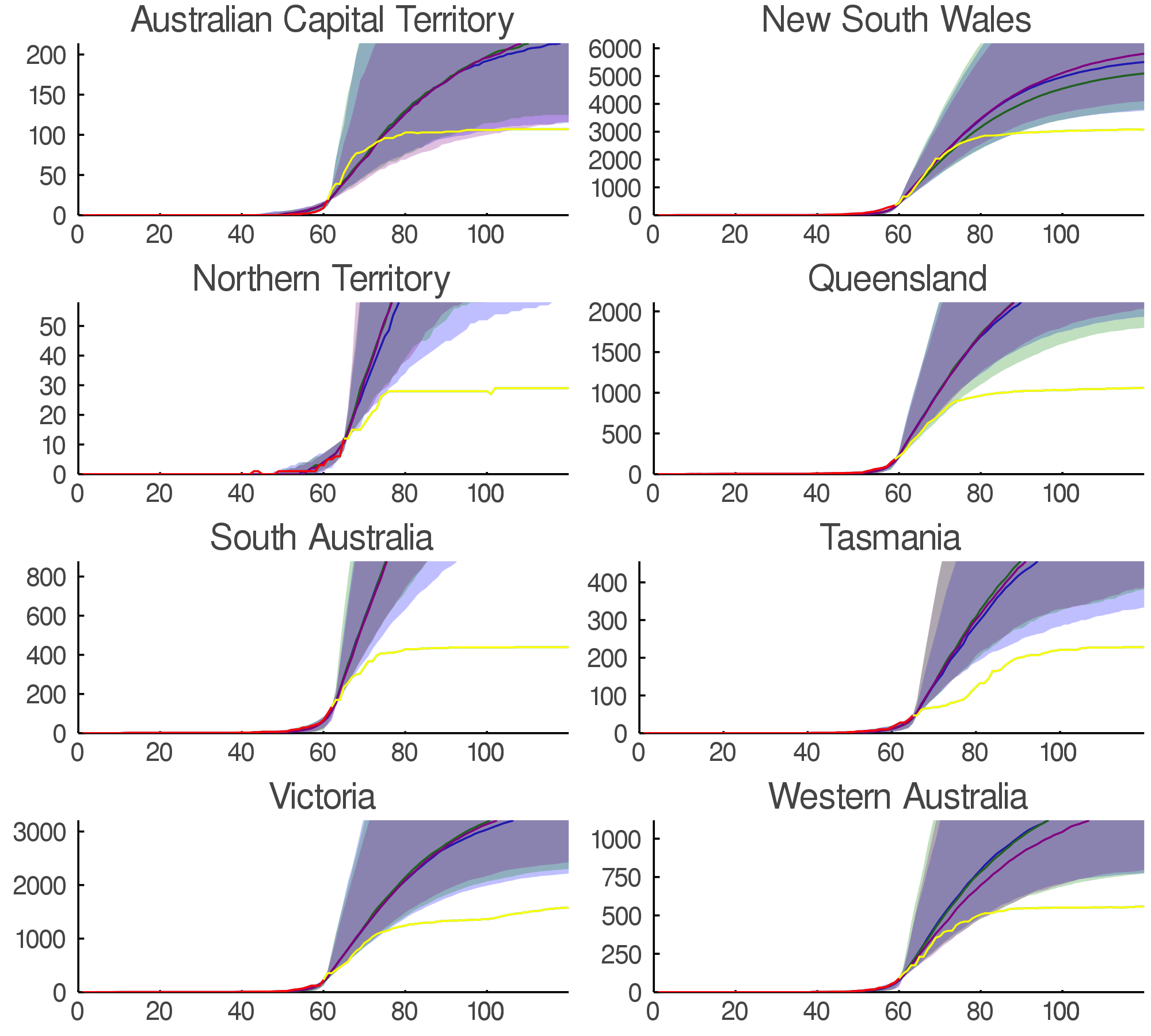}
\caption{Control evaluation. We depict the effectiveness of control measures for  \red{each Australian state and internal territory (excluding Jervis Bay)}. In each case the epidemic diffusion is fitted to data up to the end of the exponential growth phase (that is, the point of inflexion on curves $S(0)-S(t)$). Simulations up to this time point $t^*$ effectively seed the network and provide a distribution of infectious and exposed individuals within the community. Beyond this point we simulate the application of small-world control network structure $L(s)$ for various values of $s$. Here we illustrate $s=0.013$, $s=0.026$ and $s=0.054$ corresponding to $95\%$, $90\%$ and $80\%$ control. Actual observed time series data is also shown and illustrates exception effectiveness of control measures for various Australian states.}
\label{control}
\end{figure*}

A separate problem is to determine the maximum likelihood values of the parameters $p$, $q$, and $r$ for a given population $N$ and $I_\rm{th}$ from an observed time series. This can be decomposed to several discrete steps.
\begin{enumerate}
\item We suppose that $q$ is fixed and estimable by other means. For COVID-19, for example, $q$ should yield a latency period of 7-14 days \cite{gM20,fM20}. Hence $q\in(\frac{1}{14},\frac{1}{7})$.
\item Determine the epidemic peak from the time series  ---  this will define the turning point and the time when growth changes from exponential for geometric. This will allow one to determine $I_\rm{th}$ and the corresponding $t^*$. In effect we are now seeking a turning point of the total number of infections ($S(0)-S(t)$) and not just $I(t)$ as done in Fig. \ref{example}.
\item For $t<t^*$ determine $p(t<t^*)$ and $r(t<t^*)$. The ratio of these two parameters determines the epidemic growth rate via $R_0$ 
\item For $t>t^*$ it remains to determine $s$, $p(t>t^*)$ and $r(t>t^*)$. We note that $s$ controls the extent to which the system is driven by diffusion (geometric) versus exponential growth. But, for now, the best we can do is a greedy likelihood maximisation process.
\end{enumerate}

Note that, in the event that the peak has not yet been reached (i.e. $t<t^*$) it is not even sensible to attempt to estimate the parameters $s$, $p(t>t^*)$ and $r(t>t^*)$. Nonetheless, in this situation one can estimate instantaneous (or windowed) values for $R_0$ and attempt to pick the end of the exponential growth phase. \red{The latency introduced by $q$ somewhat complicates this process.} 
Figure \ref{control} illustrates the result of such a calculation. Finally, we note (as is indicated in the illustrated exemplars) that we assume a single policy change-point $t^*$ --- this is clearly inappropriate for more complex time dependent responses.\footnote{Or for regimes with inconsistent, indecisive or ineffectual responses.} \red{Of course the value of $t^*$ is actually determined by societal responses and control measures instituted in response to an outbreak. That is, it should, in principal be observable. Nonetheless, it is not clear that this will necessarily translate to the time when control takes effect --- nor will it necessarily be possible to reduce it to a single control point. Hence, the value of $t^*$ we introduce here is a single parameter value corresponding to the single moment in time when a broad range of control measures modify the dynamics of the epidemic.}

\subsection{\red{Controlled Recovery}}
\label{recovery}

\begin{figure*}[t!]
\centering
\includegraphics[width=0.8\linewidth]{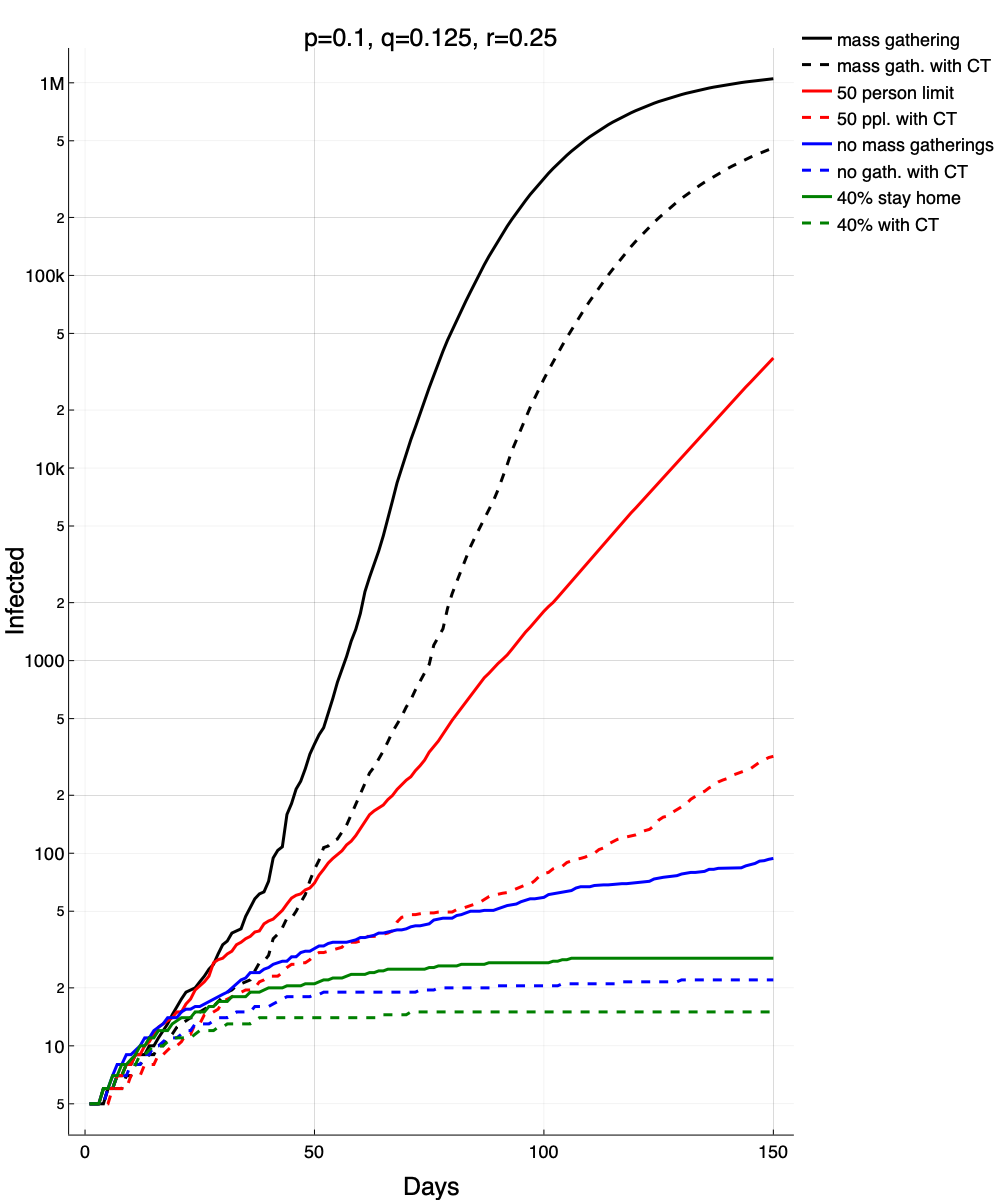}
\caption{\red{Recovery and return. Here we depict the effect of various palliative control measures in the event of a reemergence of infection (modelled here by a population seeded with $5$ exposed (infected but asymptomatic) individuals. The four solid lines represent a return to mass gatherings (black), a 50 person limit on gatherings (red), no mass gatherings (blue), and continued physical distancing (green). The dashed lines model the same scenarios with the addition of $50\%$ of the population adopting and using contact tracing software (CT). Note that the red (second solid) line grows exponentially, the black line (top) is faster than exponential and the blue and green (bottom) lines are significantly below exponential. In all cases these lines represent the median of $100$ simulations.}}
\label{relapse}
\end{figure*}

\red{Finally, in Fig. \ref{relapse} we explore the effect of control measures to mitigate against reemergence of the virus. We assume a healthy population and five individuals in state $E$. We then simulate various different control measures, again modelled via complex networks as contact graphs. The population is $2.1\times 10^6$, as before, and the parameters $p=\frac{1}{10}$, $q=\frac{1}{8}$ and $r=\frac{1}{4}$ represent a state of heightened vigilance --- but not sufficient to suppress infection. Each of the control measures described in the figure is modelled as follows
\begin{itemize}
   \item {\bf Mass gatherings} are modelled with $A=B(k)$ a scale-free network and hence no upper bound on the number of contacts a node might have.
    \item {\bf Contact tracing (CT)} is modelled by assuming that a fraction $w$ of the population has adopted contact tracing through their mobile device. Hence, if an infection were to occur between two such individuals, that infection will be extinguished via intervention from authorities. The fraction of links that are effectively removed is $w^2$.
    \item {\bf{$N$ person limit}} is modelled by truncating the scale-free network so that no node has degree larger than $N$. This is equivalent to the treatment described in \cite{xF08}.
    \item {\bf No Mass gatherings} are modelled with $A=L(1)$, as before. 
    \item {\bf{$d\%$ Stay home}} models physical isolation of a fraction $\frac{d}{100}$ of the population and is modelled with a small world lattice $L(s)$ where $d=(1-s)^k$.
\end{itemize}}

\section {Social Context and Utility}
\label{social}

\begin{figure*}[t!]
\centering
\includegraphics[width=\linewidth]{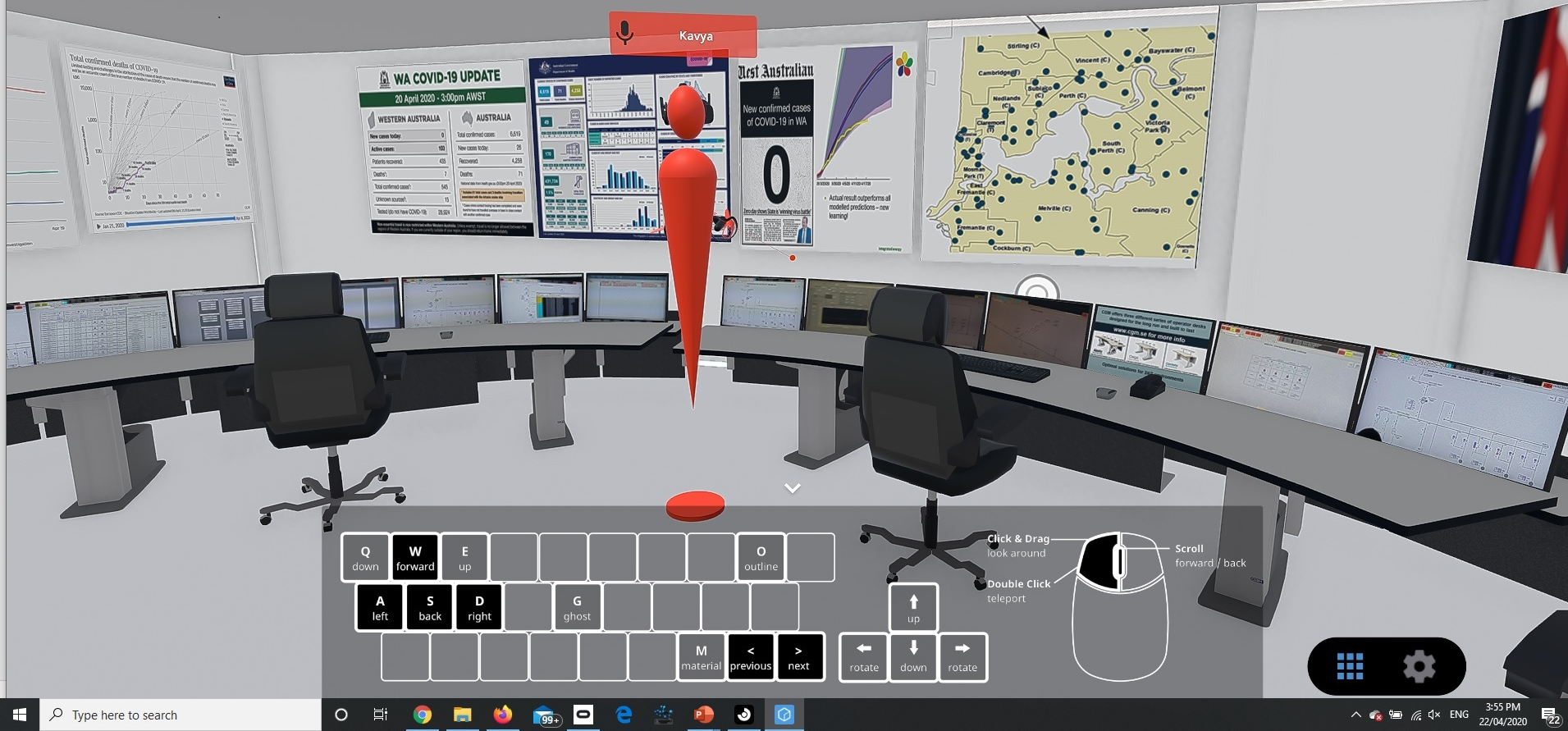}
\caption{Immersive multiperson visualisation of the model in context using virtual reality.  Here illustrated in our implementation of  a digital Public Health Emergency Operations Centre, where the model is integrated into wider contextual information such as national trends and geospatial information.  The model is used to communicate scenarios allowing stakeholders able to draw conclusions collaboratively in context.}
\label{pheoc}
\end{figure*}

This model has its origins in the severe societal challenge of COVID-19, when the population of Perth was facing the prospect of loss of 30,000 lives, and hospitals being over-run within two or three weeks if the rate of escalation continued.  The model was first used in a pandemic response workshop for a city of 100,000 people, \red{led by the second author.} \red{The model results informed the importance in influencing people's behaviour, to greater than $90\%$  compliance, and hence the guidance to give to the city officers in the workshop.} \red{It} served to demonstrate the dramatic range of outcomes which were possible, depending on the behaviour of constituents of the city, and degree of social distancing achieved.  This proved very effective in enabling appropriate action, both in the workshop and afterwards \red{with the city response seen as a model}.  Subsequently the results were shared on professional social media, and an online conference, influencing thousands more.

In combination with effective timely coordinated state and federal government polices, and a high level of societal compliance, a very strong result of virus suppression was achieved.
The model was further developed to update progress, within two weeks, and at the time of the workshop debrief this was used to show the importance of continuing measures in suppression, \red{and} the rate at which rapid outbreak could occur, even in the context of strong initial suppression.  This allowed the appropriate focus to shift towards a positive recovery.  \red{Again} this was shared local\red{ly} and internationally to provide hope for others and influence behaviour.

Subsequently, the actual case data within the state was plotted against the forecast range, and this was shared with state scientific authorities, enabling a constructive discussion about the correlation between application of selected state and national control measures and outcomes.  The extension to modelling different approaches to recovery continues in a similar mode, \red{with distinctive results, and the model outcomes have since been included in briefings for state health authorities and COVID-19 safety training.}

To gain most value from the model, its results have been interpreted in a variety of environments, including most recently in collaborative virtual reality mode, in a digital Public Health Emergency Operations Centre (PHEOC) (Fig. \ref{pheoc}).  This has the advantage of rich immersion in the data, while allowing deep multi-party interaction and dialogue to discern appropriate observations, and at the same time allow parties to engage together from anywhere in the world.  At the time of drafting, the number of new cases of COVID-19 has for the first time reached zero, with only seven fatalities in the State to date, remarkably low compared to world averages.  \red{A few weeks later the disease had been eliminated from the hospital system in Western Australia.}

\section{Conclusion}

The model we present here has a small -- perhaps minimal -- number of parameters, and describes the observed dynamics of pandemic disease transmission. When applied to data from the global outbreak of \red{coronavirus} in 2019/2020, the model \red{provides good qualitative agreement with} observed data across population centres. Nonetheless, identical simulations with new initial conditions yield vastly different outcomes. The variance of our model predictions is large, and \red{in fact much larger than the variance observed between distinct epidemiological parameter values. Hence, choice of {\em optimal} transmission rates is a secondary concern behind appropriate description of contact patterns and transmission mitigation strategy.} Our results indicate that particular simulations of models that claim to have predictive power within that prediction envelope may be prone to over-interpretation. Finally, despite modelling a \red{complex system with} complex networks we have demonstrated the sufficiency of a minimal model. Models with large numbers of parameters {\em which are fitted to time series data} are unnecessary and likely to be unreliable and misrepresent the underlying dynamical process. \red{Our model emphasises accurate reproduction of the qualitative behaviour of the system, this} does not preclude the construction of more complex models when sound epidemiological reasoning dictates it is necessary and when informed with direct evidence to allow for quantitative estimation of the relevant parameters.

While we are reluctant to make predictions from, or over interpret the application of, this model to the current coronovirus pandemic, our results indicate that strict physical isolation in combination with monitoring and the usual transmission mitigation strategies are required to minimise impact. Below $80\%$  compliance with physical isolation measures risks catastrophic spread of infection (Fig. \ref{parameters2}). This data is consistent with the evidence of explosive growth of infection experienced in some localities. Without decisive and potentially severe intervention, similar disasters are likely to occur in regions with weaker health systems.

In the simulations described above, we do not make any attempt to ensure ``pseudo-continuity'' between time varying manifestations of $A$. That is, nodes that are connected for one network are not more likely to be connected after switching the network topology. We could see no simple and generic way in which to achieve this. Moreover, we did not detect any excessive mixing that one might expect should this mismatch be an issue.

\red{It is worth noting that the computation cost of this model --- despite being a population level simulation --- is not great. We simulate the state of an entire population, but at each iteration the updates are determined entirely by a predefined contact structure. For population size $N$ and simulating $T$ time steps the computational time cost is $NT$. The memory requirement is $N+N\log{N}$ (for a sparse contact network and population state vector). This modest computation demand mean that the algorithm can be successfully deployed in immersive, interactive and real-time environments.}

\red{In Sec. } \ref{state} we raise the issue of estimating the expected distribution of unobserved infection sites (i.e. state $E$) on a network. Should the model described here prove relevant, this will be an issue of immense importance to the proper quantification of uncertain future behaviour.  \red{Figure \ref{relapse} illustrates the application of these technique for future scenario planning.}

Finally, the social context and utility of this modelling is demonstrated by its live use in shaping the planning and implementation of a highly effective response to COVID-19 on a city and state level. Ultimately, one must ask what is the purpose of modelling. Epidemic disease transmission is a fairly simple mathematical problem --- exponential growth followed by decay. The difficulty is in reliably estimating parameters. We show that the contact structure provides a direct and effective approach to model control strategies. In addition to the information provided by our simulations, we describe in Sec. \ref{social} the application of these methods to effectively inform and influence policy makers.
\\

\section*{Data Availability}

Source code for all calculations described in the manuscript is available on \url{https://github.com/m-small/epinets}. Data was obtained from \url{https://github.com/CSSEGISandData/COVID-19}.



\bibliography{bibliography}

\begin{IEEEbiography}[{\includegraphics[width=1in,height=1.25in,clip,keepaspectratio]{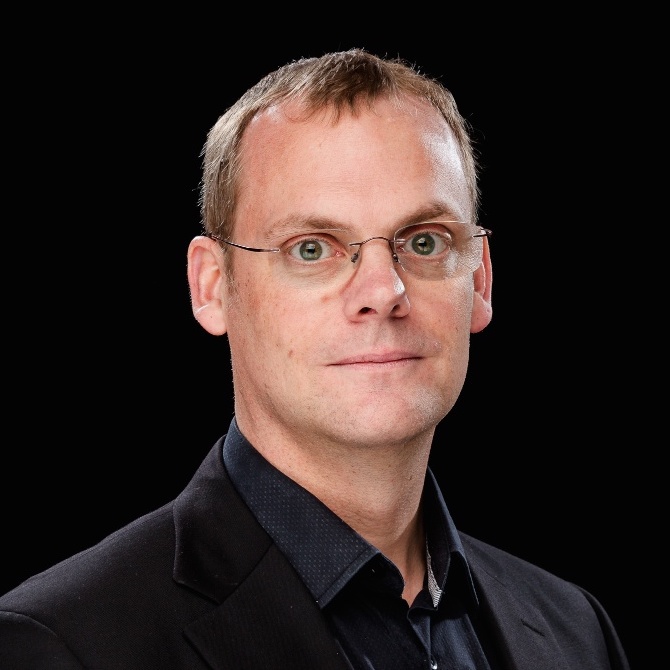}}]{Michael Small}  (M'01--SM'08) received undergraduate and doctoral degrees in pure and applied mathematics from the University of Western Australia (UWA), Perth, WA, Australia. After postdoctoral experience in Stellenbosch and Edinburgh, he joined the faculty of the Department of Electronic and Information Engineering, Hong Kong Polytechnic University (2001-2011). In 2011, he was the recipient of Australian Research Council Future Fellowship and in 2012 was made Winthrop Professor of Applied Mathematics in the School of Mathematics and Statistics, UWA. Since 2015 he has held the UWA-CSIRO Chair of Complex Engineering Systems. He is Editor of the journal {\em Chaos} and Associate Editor of {\em International Journal of Bifurcations and Chaos}. His research interests include complex systems, complex networks, chaos and nonlinear dynamics, nonlinear time series analysis, and computational modelling. Michael is Principal Modelling Consultant with Integrated Energy Pty Ltd.
\end{IEEEbiography}

\begin{IEEEbiography}[{\includegraphics[width=1in,height=1.25in,clip,keepaspectratio]{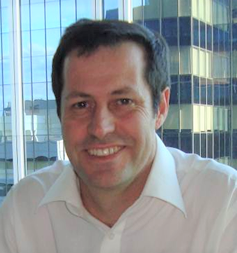}}]{David Cavanagh} studied Medicine and Engineering at the University of Western Australia.  He is Managing Director of Integrated Energy Pty Ltd, who have been working for the last decade to enable safer, more valuable companies and communities, which are better for the environment and people.  David leads a multidisciplinary consulting team which spans all the timezones of the world, with a proven track record in consulting assessments which have saved and enhanced lives in addition to tens of billions of dollars of value for his clients globally.  David brings three decades of experience, with locations spanning six continents, and into space.  His practice brings together psychology, technology and architecture, so people can work better together over distance.  A worldwide COVID knowledge network supports client engagements at company, city, state and national level.
A global client list includes City Councils, Universities, Australian State and Federal Government Agencies, many of the world’s leading energy and resources companies, technology companies such as IBM; clean energy startups, and the European Space Agency.  He is the author of several publications including a book for Shell on how society should manage Y2K, wrote and delivers a course which has been delivered in six countries and three languages, chairs global conference streams in Health, Safety and Environment, and has won national innovation awards.

\end{IEEEbiography}

\EOD

\end{document}